\documentclass{PoS}
\usepackage{epsfig}
\PoS{PoS(LAT2005)212}
\newcommand{\bea}{\begin{eqnarray}}
\newcommand{\eea}{\end{eqnarray}}
\newcommand{\nn}{\nonumber}

\title{Pionic couplings \boldmath{$\widehat g$} and \boldmath{$\widetilde g$}\\ 
in the static heavy quark limit}

\ShortTitle{Pionic couplings $\widehat g$ and $\widetilde g$ in the static heavy quark limit}

\author{D.~Be\'cirevi\'c, B.~Blossier, Ph.~Boucaud, J.P.~Leroy, A.~LeYaouanc and O.~P\`ene\\
Laboratoire de Physique Th\'eorique (B\^at.210), Universit\'e Paris Sud, Centre d'Orsay,\\ 
F-91405 Orsay Cedex, France\\
E-mail: \email{Damir.Becirevic@th.u-psud.fr}
}

\abstract{The couplings between the soft pion and the doublet of heavy-light mesons
are basic parameters of the ChPT approach to the heavy-light systems. We compute 
the unquenched ($N_f=2$) values of such couplings in the static heavy quark limit: 
(1) $\widehat g$, coupling to the lowest doublet of heavy-light mesons, and 
(2) $\widetilde g$, coupling to the first orbital excitations. A brief description 
of the calculation together with the short discussion of the results is presented.}

\FullConference{XXIIIrd International Symposium on Lattice Field Theory\\

                 25-30 July 2005\\

                 Trinity College, Dublin, Ireland}

\begin{document}

\section{Physics Motivation}
Static quark limit of QCD offers a simplified framework to solving the non-perturbative dynamics 
of light degrees of freedom, which is highly important for various phenomenological studies of 
weak interactions involving the heavy-light mesons. In the exact $m_Q\to \infty$ limit, 
the heavy quark symmetry (HQS) constrains the QCD dynamics of light quarks and soft gluons in the heavy-light hadrons 
to be invariant under the change of the heavy quark flavour and/or its spin ($j_Q$). 
As a result the total angular momentum of the light degrees of freedom becomes a good quantum number 
($j_\ell^P$), and therefore the physical heavy-light mesons come in mass-degenerate doublets:
\bea
\underbrace{[D_{}(0^-), D_{}^\ast(1^-)]}_{j_\ell^P=\frac{1}{2}^- (L=0)}\,,\quad
\underbrace{[D_{0}^\ast(0^+), D_{1}^\prime(1^+)]}_{j_\ell^P=\frac{1}{2}^+ (L=1)}\,,\quad
\underbrace{[D_{1}(1^+), D_{2}^\ast(2^+)]}_{j_\ell^P=\frac{3}{2}^+ (L=1)}\,,\quad \dots
\eea
where on the example of charmed states we remind the reader of spectroscopic labels. 
HQS is a basis for the systematic expansion of the QCD Green functions in power 
series of $\Lambda_{\rm QCD}/m_Q$, known as the heavy quark effective theory (HQET). 
At each order, one has to solve the non-perturbative QCD dynamics of the light 
degrees of freedom which can actually be done only by the numerical QCD simulations 
on the lattice. 
\EPSFIGURE[r]{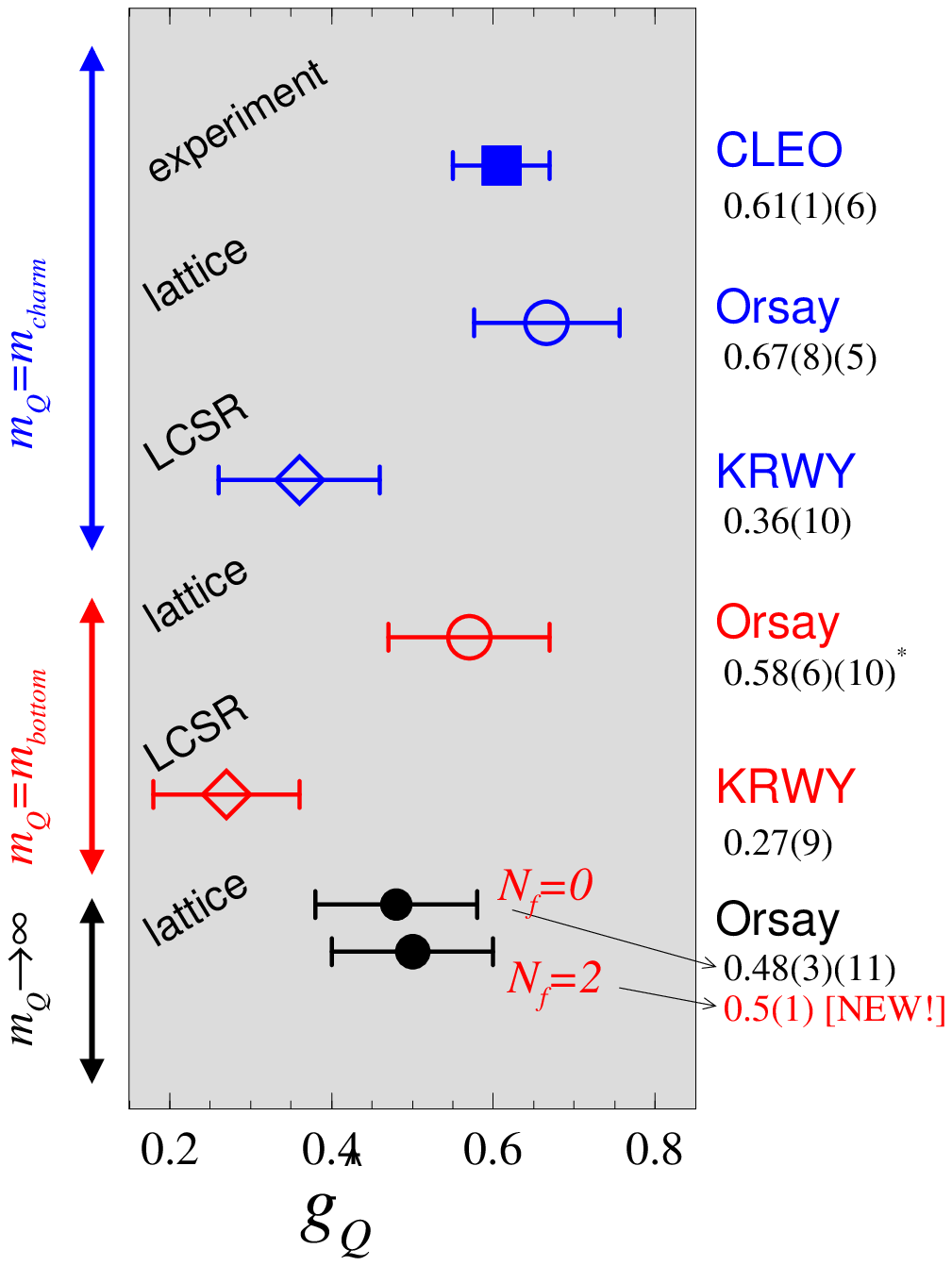,height=8cm} {\label{fig:g}
Results for $\widehat g$ with the heavy quark being static, bottom or charm. Asterisk 
on the lattice result for $\widehat g_b$ indicates that the result is not obtained after interrpolating between the static and charm heavy quark
cases. The numbers are taken from refs.~\cite{CLEO,gDDpi,KRWY,g-static}, in descending order.}
Due to inability to simulate directly the dynamics of light quarks 
that are close enough in mass to the physical $u$- or $d$-quark, one has to make 
various chiral extrapolations, which induce large systematic errors to the lattice results 
(a typical example is the one of $SU(3)$ flavour breaking ratio of the pseudoscalar 
meson decay constants, $f_{B_s}/f_{B_d}$). In addition the insufficient size of the lattice box 
with the periodic boundary conditions afflicts the propagation of the light pions on the lattice. 
To have a good handle on these two problems one can rely on the effective theory that 
describes the dynamics of light degrees of freedom in terms of pseudo-Goldstone bosons (pions,, for short), 
usually referred to as heavy meson chiral perturbation theory (HMChPT)~\cite{Casalbuoni}.  
HMChPT is suitable when working with (nearly) massless quarks and thus it is complementary 
to the light quark dynamics that is directly accessible from the lattice. 
Similar to the standard ChPT, where the pion axial coupling ($f_\pi$) is a parameter of the theory, 
in the HQChPT Lagrangians the axial couplings of a pion to doublets of heavy-light mesons are parameters of the theory. 
In other words, they have to be fixed elsewhere. 
The most well-known such a coupling is the one with the $(1/2)^-$-doublet, known as $\widehat g$, whose value 
was experimentally established in the case of heavy charm quark~\cite{CLEO}. That value appeared to be much 
larger than the ones predicted by most of the relativistic quark models and by all the QCD sum rules. 
In fig.~\ref{fig:g}, we show the panorama of the results for that coupling obtained 
by using lattices and light cone QCD sum rules (LCSR). Lattice values are obtained both in the static heavy quark limit~\cite{g-static,g-ukqcd} 
and with the heavy quark of the mass around the physical charm quark mass~\cite{gDDpi}. However, all the lattice computations 
of that coupling are obtained in the quenched approximation. In this note, we present the first unquenched result 
for $\widehat g$, with $N_f=2$ dynamical light quarks and in the static heavy quark limit (indicated as "New" in fig.~\ref{fig:g}. In addition, we report the first
result for $\widetilde g$, the coupling of the soft pion to the $(1/2)^+$-doublet of heavy-light mesons.

\section{Definitions and Correlation functions to be computed}

In the limit in which the heavy quark is infinitely heavy and the light quarks are massless, 
the axial couplings of the charged pion to the heavy-light mesons, $\widehat g$ and $\widetilde g$, are defined via
\bea
\langle H\vert \vec {\cal A}\vert H^\ast(\varepsilon) \rangle = \widehat g\ \vec \varepsilon\,, \qquad
\langle H^\ast_0\vert \vec {\cal A}\vert H^\prime_1(\varepsilon) \rangle = \widetilde g\ \vec \varepsilon\,,
\eea
where the non-relativistic normalisation of states $\vert H^{(\ast,\prime)}_{0,1}\rangle$ is assumed,  
$  \langle H_i(v)\vert H_j(v^\prime)\rangle = \delta_{ij}\delta(v - v^\prime)$. All heavy-light hadrons are at rest 
($\vec v=\vec v^\prime =\vec 0$), and therefore the soft pion which couples to the axial current,  
${\cal A}_\mu = \bar u\gamma_\mu \gamma_5 d$, is also at rest, $\vec q=\vec 0$. $\varepsilon_\mu$ is 
the polarisation of the vector and the axial heavy-light meson respectively. 

The standard strategy to compute the above matrix elements consists in evaluating the following two- 
and three- point correlation functions:
\bea\label{eq:correlators}
C_2(t) &=&\langle\sum_{\vec x} P(x) P^\dagger(0)\rangle_{_U} =\frac{1}{3}\langle\sum_{i,\vec x} V_i(x) V_i^\dagger(0)\rangle_{_U}
 =
\langle\sum_{\vec x} {\rm Tr}\left[ {1+\gamma_0\over 2}P_x^0 \gamma_5 {\cal S}_{u,d}(0,x) \gamma_5\right]\rangle_{_U} , 
\nn\\
\widetilde C_2(t) &=&\langle\sum_{\vec x} S(x) S^\dagger(0)\rangle_{_U} =\frac{1}{3}\langle\sum_{i,\vec x} A_i(x) A_i^\dagger(0)\rangle_{_U}
 =
\langle\sum_{\vec x} {\rm Tr}\left[ {1+\gamma_0\over 2}P_x^0 {\cal S}_{u,d}(0,x)\right]\rangle_{_U}  ,
\nn\\
C_3(t_x)&=&\langle\sum_{\vec x,\vec y} P(y) {\cal A}_i(x) V_i^\dagger(0)\rangle =
\langle\sum_{\vec x,\vec y} {\rm Tr}\left[ {1+\gamma_0\over 2}P_y^0 \gamma_i {\cal S}_u(0,x) \gamma_i\gamma_5
{\cal S}_d(x,y)\gamma_5\right]\rangle_{_U}  ,
\\
\widetilde C_3(t_x)&=&\langle\sum_{\vec x,\vec y} S(y) {\cal A}_i(x) A_i^\dagger(0)\rangle_{_U} =
\langle\sum_{\vec x,\vec y} {\rm Tr}\left[ {1+\gamma_0\over 2}P_y^0 \gamma_i \gamma_5 {\cal S}_u(0,x) \gamma_i\gamma_5
{\cal S}_d(x,y)\right]\rangle_{_U}\nn  ,
\eea
where $\langle\dots\rangle_{_U}$ denotes the average over independent gauge field configurations. 
The gauge field configurations used in this work are the ones produced by the SPQcdR collaboration~\cite{spqcdr}. 
The interpolating fields that we use to produce the desired $(1/2)^\mp$ heavy-light mesons are: 
$P= \bar h \gamma_5 q$, $V_i= \bar h \gamma_i q$, $S= \bar h q$, and $A_i= \bar h \gamma_i \gamma_5 q$, 
with $q$ being either $u$- or $d$-quark field.  In eq.~(\ref{eq:correlators}) 
we also express the correlation functions in terms of quark propagators: light ones, ${\cal S}_{u,d}(0,x)$,
and the infinitely heavy one (the Wilson line), 
\bea\label{Pline}
P_x^0 = \delta(\vec x) \prod_{\tau=0}^{t_x-1}U_0(\tau)\,,
\eea
where $U_0$ is the temporal component of the link variable. 
The spectral decomposition of the three point function $C_3(t_x)$ reads
\bea\label{eqC3}
C_3(t_x) = \sum_{m,n} \Big[ \underbrace{\langle 0\vert P\vert H_n\rangle}_{Z_n}e^{-E_{H_n}t_y} \langle H_n\vert {\cal A}_i\vert H^\ast_m\rangle 
e^{-(E_{H_m^\ast}-E_{H_n})t_x} 
\underbrace{\langle H^\ast_m\vert V_i\vert 0\rangle}_{Z_m \varepsilon_i(m)} \Big]\,,
\eea 
with the sum running over the ground states ($m=n=0$) and the radial excitations 
($m,n>0$) which are heavier and thus exponentially suppressed. If only 
the diagonal terms were different from zero [i.e., the terms in the
sum~(\ref{eqC3}) with $m=n$], 
then the function $C_3(t_x)$ would be $t_x$-independent because the HQS ensures 
that $E_{H_n^\ast}=E_{H_n}$. This is indeed what one observes from the lattice 
data: even for very small values of $t_x$ 
the correlation functions $C_3(t_x)$ and  $\widetilde C_3(t_x)$ exhibit 
plateaus (c.f. fig.~\ref{fig:plateaus}). 
However, even after leaving out the non-diagonal terms ($m\neq n$), 
we still have a problem of contamination by the radial excitations 
($m=n>0$) in  eq.~(\ref{eqC3}). The latter are taken care of by 
implementing the smearing procedure which highly enhances the overlap of the interpolating fields with   
the lowest energy states. Once that is been done, it is then trivial to extract the couplings $\widehat g$ and $\widetilde g$, namely 
\bea
{C_3(t_x)\over (Z_0^{\rm sm.})^2 e^{-E_{0}t_y}}\longrightarrow \widehat g(m_\pi^2)\,,\qquad
{\widetilde C_3(t_x)\over (\widetilde Z_0^{\rm sm.})^2 e^{-\widetilde E_{0}t_y}}\longrightarrow \widetilde  g(m_\pi^2)\,.
\eea
where $(Z_0^{\rm sm.})^2$ is extracted from the fit to $C_2(t)= \displaystyle{\sum_n} |Z_n^{\rm sm.}|^2 \exp(-E_n t)$, with smeared sources. 
Likewise  $(\widetilde Z_0^{\rm sm.})^2$ is extracted from $\widetilde C_2(t)$. Before going to the results, we should stress again that the above identification with the
couplings $\widehat g$ and $\widetilde g$ is valid only in the chiral limit, $g=\displaystyle{\lim_{m_\pi^2\to 0}g(m_\pi^2)}$. 
\EPSFIGURE[l]{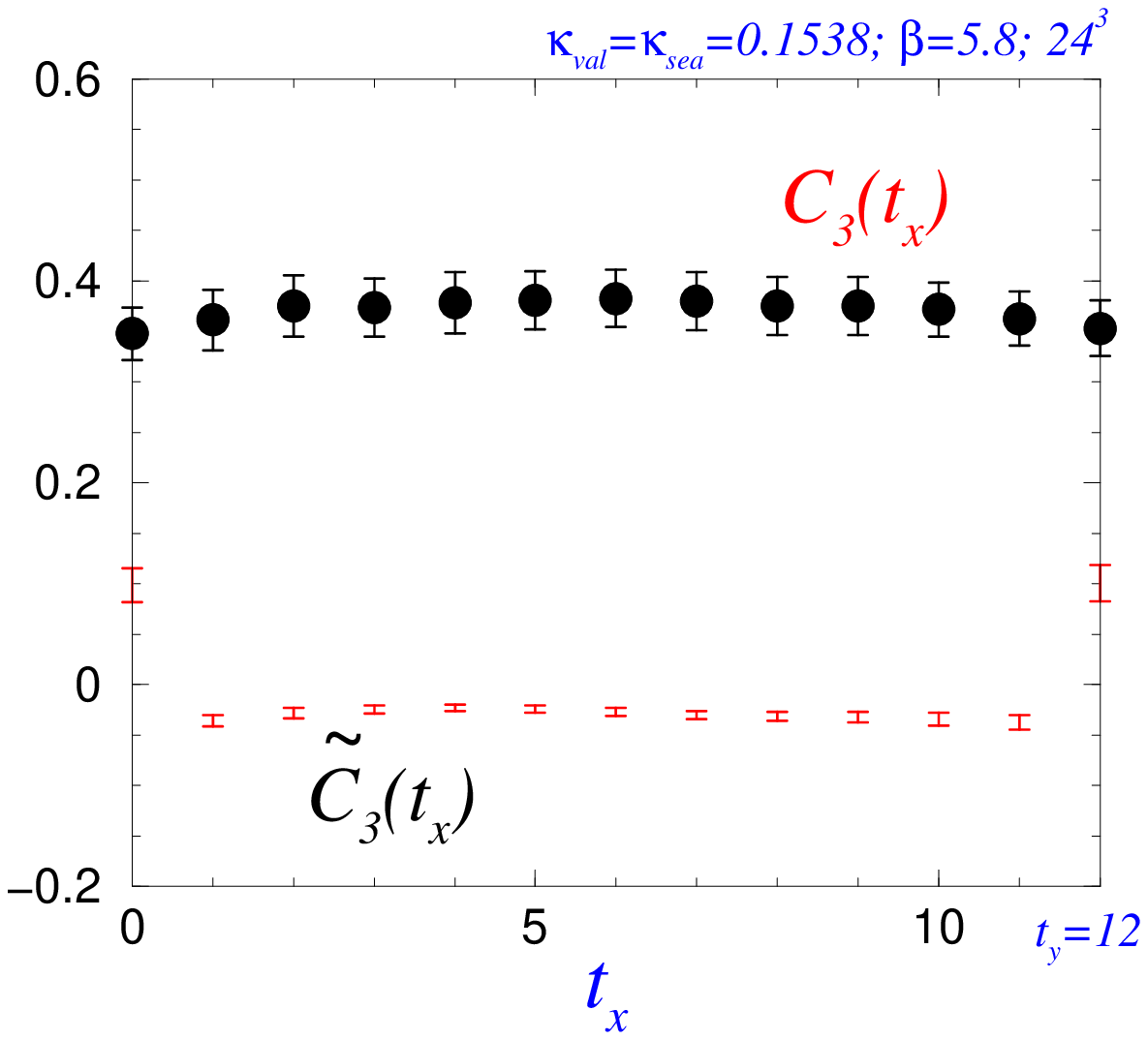,width=7.4cm} {\label{fig:plateaus}
Correlation functions $C_3(t_x)$ and $\widetilde C_3(t_x)$ in which the sources are fixed at $0$ and 
$t_y=12$.}

\section{Lattice details and results}

Our dataset consists of $50$ gauge field configurations which include $N_f=2$ dynamical quarks whose masses, 
with respect to the physical strange quark mass, are within the range $0.7\lesssim m_q/m_s^{phys}\lesssim 1.3$. 
We use the (unimproved) Wilson quarks and work on $24^3\times 48$ lattices at $\beta=5.8$, 
which corresponds to the lattice spacing of $a\approx 0.06$~fm, and the physical volume $(1.5~{\rm fm})^3$. 
The corresponding "sea" quark hopping parameters ($\kappa_{\rm sea}$) are listed in table~\ref{tab:res}. 
More information about the simulations can be found in ref.~\cite{spqcdr}. In all our computations we keep the 
valence and sea light quarks equal.

In the computation we used the Eichten--Hill static quark action~\cite{Eichten-Hill}, supplemented 
by the hypercubic blocking procedure (HYP) of ref.~\cite{HYP} which is known to substantially improve 
the signal/noise ratio in the correlation functions~(\ref{eq:correlators}). In that way we were able to monitor the 
the efficiency of the smearing, i.e., by confronting the signal of $C_2(t)$ by using both local sources with the signal 
in which both sources are being smeared. The generic local source operator is smeared as in ref.~\cite{Boyle}, $\bar h(x) \Gamma q(x) \to 
\bar h(x) \Gamma q^S(x)$ where, 
\bea \label{Smearing}
q^S(x)= \sum\limits_{r=0}^{R_{\rm max}} {\scriptstyle
(r+\frac{1}{2})}^2 e^{-r/R}
 \sum\limits_{k=x,y,z} \left[
q({\scriptstyle x+r\hat{k}})\times \prod\limits_{i=1}^r U^F_k({\scriptstyle
x+(i-1)\hat{k}})  + q({\scriptstyle x-r \hat{k}})\times
\prod\limits_{i=1}^r U_k^{F^\dagger}({\scriptstyle
x-i\hat{k}})\right]  \,,
\eea 
$U^F$ being the so called fuzzed link variable. The parameters $R=3$ and $R_{\rm max}=5$ are 
tuned to enhance the overlap with the lowest lying state, which we were able to check from the 
analysis of the two point functions $C_2(t)$ and $\widetilde C_2(t)$. We obtain that 
$Z_0^{\rm sm.}/Z_0^{\rm loc.}\approx 150$, while $Z_1^{\rm sm.}/Z_1^{\rm loc.} < 0.05$. 
Similar situation holds for the scalar-scalar correlation function, and it is therefore safe to say that 
the pollution of our results for $\widehat g(m_\pi^2)$ and $\widetilde g(m_\pi^2)$ that comes from 
the transitions among higher excitations is to a huge extent suppressed by the smearing procedure.
\begin{table}[h]
\begin{center}
\begin{tabular}{|c|cccc|}\hline
$\kappa_{\rm sea}=\kappa_{\rm val}$ & $r_0 m_\pi$ & $r_0(\widetilde E_0-E_0)$  & $\widehat g(m_\pi^2)$ & $\widetilde g(m_\pi^2)$ \\ \hline
{\phantom{\huge{l}}} \raisebox{-.2cm} {\phantom{\huge{j}}}
$0.1535$ & $1.97(5)$    & $1.49(11)$  & $0.55(4)$ & $-0.18(1)$	\\
{\phantom{\huge{l}}} \raisebox{-.2cm} {\phantom{\huge{j}}}
$0.1538$ & $1.82(4)$    & $1.71(12)$  & $0.48(3)$& $-0.20(4)$	\\
{\phantom{\huge{l}}} \raisebox{-.2cm} {\phantom{\huge{j}}}
$0.1540$ & $1.72(5)$    & $1.60(12)$  & $0.52(6)$& $-0.21(4)$	\\
{\phantom{\huge{l}}} \raisebox{-.2cm} {\phantom{\huge{j}}}
$0.1541$ & $1.47(7)$    & $1.74(19)$  & $0.43(4)$& $-0.14(2)$	\\ \hline
\end{tabular}\caption{\label{tab:res}Results for the orbital splitting $\widetilde E_0-E_0$, the couplings 
$\widehat g$ and $\widetilde g$, from the unquenched $N_f=2$ simulations in which the dynamical quark is of  
the mass corresponding to the pion whose masses are also given.}
\end{center}
\vspace*{-3mm}\end{table}\\

\begin{figure}[b]
\begin{center}
\includegraphics[width=7.27cm,clip]{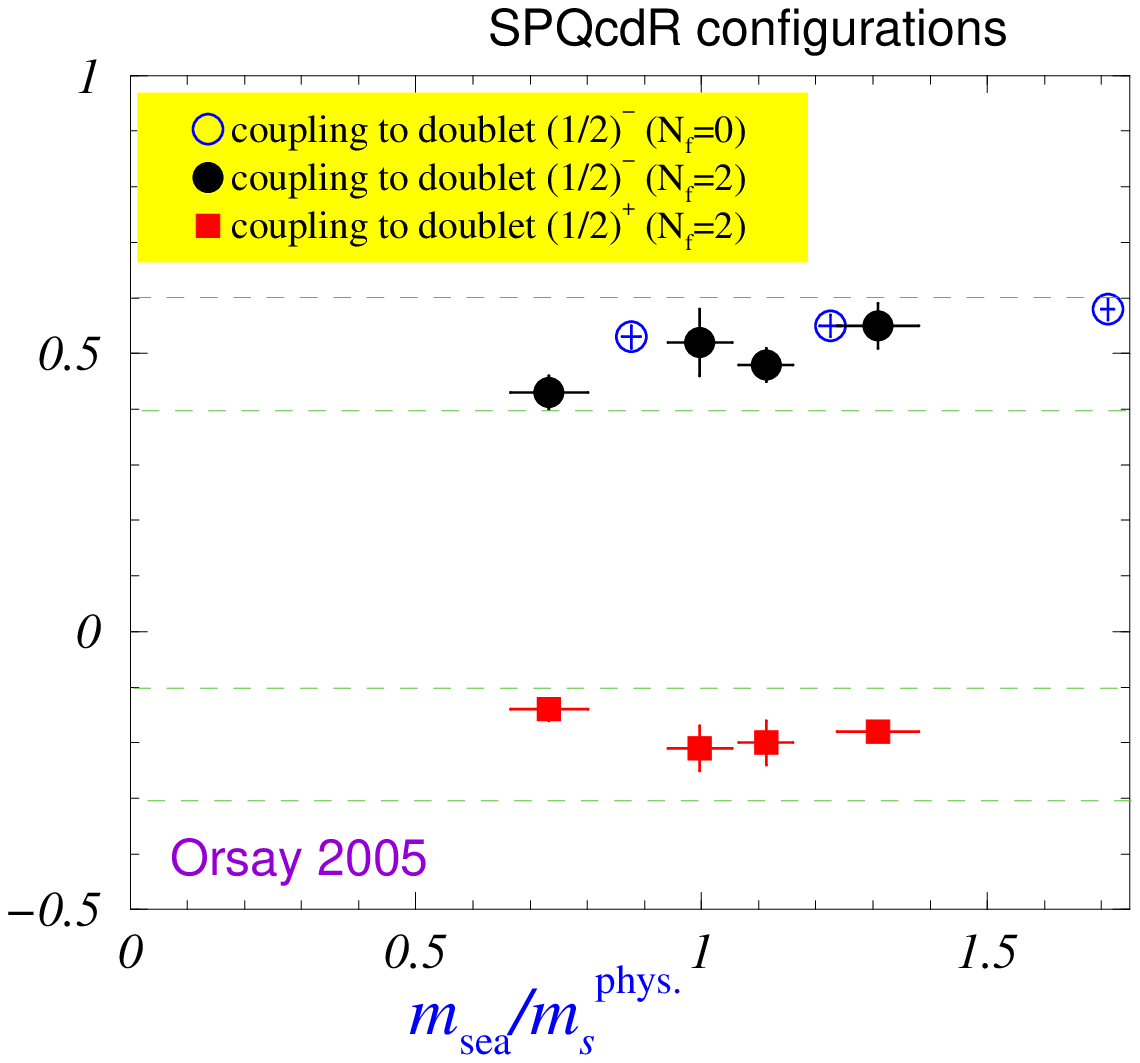}
\caption{\label{plot2} Our new (unquenched) results against the old (quenched) ones. 
The dashed lines correspond to the range of values that $\widehat g(0)$ and $\widetilde g(0)$ can assume 
after the chiral extrapolation.}
\end{center}
\end{figure}

Our results for $\widehat g (m_\pi^2)$ and $\widetilde g (m_\pi^2)$ are given in table~\ref{tab:res}, 
and plotted in fig.~\ref{plot2}. These results are obtained by fixing one 
of the source operators to $t_y=12$. The stability of these values is checked by using $t_y=11$, and $t_y=13$. In the same table we also give the mass
difference between the lowest lying states belonging to $(1/2)^+$ and $(1/2)^-$ doublets.  
On the basis of our results, we make the following observations:
\begin{itemize}
\item The values of $\widehat g(m_\pi^2)$ from the unquenched simulations are consistent with the quenched 
values we reported in ref.~\cite{g-static} (see also fig.~\ref{plot2});
\item The values of $\widetilde g(m_\pi^2)$ are clearly negative and their absolute values much smaller 
than those of $\widehat g(m_\pi^2)$;
\item The statistical and systematic quality of our data is not good enough to discuss the subtleties of 
chiral extrapolations. Instead, we quote $0.4\leq \widehat g \leq 0.6$, and $-0.3 \leq \widetilde g \leq -0.1$, 
as our current estimates;
\item Less prone to statistical noise is the ratio $\widetilde g/\widehat g$, for which we obtain
\bea
\widetilde g/\widehat g = -0.34(3),-0.43(9),-0.39(7),-0.31(6),
\eea
in the order of decreasing light quark mass; 
\item Our results are in a clear conflict with the assumption, $\widetilde g=\widehat g$, that stems from the 
parity doubling models~\cite{pdm}; 
\item Our values of the orbital splitting $\widetilde E_0 - E_0$, together with the couplings 
$\widehat g$ and $\widetilde g$ do not help to explain the puzzling experimental phenomenon that 
$(\widetilde E_0 - E_0)_s < (\widetilde E_0 - E_0)_{u,d}$~\cite{puzzle}. 
To address this question more seriously it is essential to reduce the sea quark masses, work on larger lattice volumes 
and increase the statistical quality of our data.
\end{itemize}
%%%%%%%%%%%%%%%%%%%%%%%%%%%%%%%%%%%%%%%%%%%%%%%%%%%%%%%%%%%%%%%%%%%%%%%%%%%%%%%%%%

\end{document}